# Development of a technology for manufacturing a heat-shielding structure on nitrogen cryocontainers, excluding heat transfer through gas

H. H. Zhun[1], V.V. Starikov[1] and V. P. Koverya[2]

[1]*National Technical University "Kharkiv Polytechnic Institute", Kharkiv 61002, Ukraine*
[2]*B. Verkin Institute for Low Temperature Physics and Engineering of the National Academy of Sciences of Ukraine*
*Kharkiv 61103, Ukraine*

E-mail: koverya@ilt.kharkov.ua

One of the important stages in the creation of the scientific and technical foundations for the calculation, design and manufacturing technology of the lowest heat-conductivity thermal protection from screen-vacuum thermal insulation (SVTI) is the development of a process for achieving the optimal vacuum $P_0 \leq 10^{-3}$ Pa in the SVTI layers, since at this pressure, thermal conductivity ($\lambda_{ef}$) through the SVTI is carried out only due to the radiant ($\lambda_{rad}$) and contact-conductive ($\lambda_{k.k}$) components. It is proposed to obtain such a pressure in thermal insulation by using cushioning material in it, which was previously degassed in a separate vacuum chamber at 370–380 K for 12 hours in order to remove water molecules from its structure and then replace them with dry nitrogen molecules. These molecules have 3–4 times less heat of adsorption; therefore they are pumped out faster. As a result, it becomes possible to accelerate (by ~ 20 hours) to achieve optimal vacuum in thermal insulation, as well as 11% lower effective thermal conductivity (equal to $(14.1–14.3) \cdot 10^{-5}$ W / (m · K)). The analysis carried out (according to the developed methodology) showed that the achieved optimal effective thermal conductivity of thermal insulation in a cryocontainers is determined by 33% of radiant thermal conductivity ($4.7 \cdot 10^{-5}$ W / (m · K)) and 67% of the contact-conductive component ($9.4 \cdot 10^{-5}$ W / (m · K)).

## 1. Introduction

The development of methods for long-term storage in a liquid nitrogen environment of vital donor reserves of skin, blood and various organs in cryomedicine and cryobiology, as well as genetic materials (in the latest technology of artificial insemination in animal husbandry) required the production of significant quantities of cryobiological containers with high-quality thermal protection. In addition, the development of a technology for the manufacture of highly efficient thermal protection at the nitrogen level of temperatures is also relevant for devices and technologies based on high-temperature superconductivity (HTSC). Their development has become possible as a result of more than 30 years of research on the mechanisms of this superconductivity by scientists from many countries, including Ukraine [1–12]. Effective thermal protection is also required at the helium temperature level for devices using low-temperature supercon-ductivity (LTSC) [13–16]. However, nitrogen cryogenic containers manufactured according to empirical technologies were of poor quality, despite the use of the lowest heat-conductivity screen-vacuum thermal insulation SVTI in their heat shield. These containers were characterized by large losses of liquid nitrogen. The reasons for this were incomprehensible. In addition, the theoretical as well as the experimental methods of their study remained unknown. Therefore, for a long time, both foreign and domestic cryobiological and other cryogenic containers, manufactured by many thousands, have not improved.

The first studies of the reasons for the low quality of nitrogen cryobiological containers, mass-produced at the Kharkov Transport Equipment Plant, were carried out by Professor Zhun H.H. at the Department of "Technical Cryophysics" of the National Technical University "KhPI". For this, the necessary computational and experimental methods were developed and experimental installations were manufactured.

At the initial stage of investigations [17] of heat inflows for all structural elements, distribution of temperature and pressure over the average integral thickness of the SVTI layers on cryocontainers with low quality, it was established that these layers were glued with the adhesive base contained in the SVTI–7 cushioning glass veil used in thermal insulation. This was a consequence of the heating of cryogenic containers to 380–390 K in electric furnaces during the implementation of the technological process of thermal vacuum degassing of their thermal insulation. As a result, local areas with a gas pressure increased by 4 orders of magnitude were formed in this thermal insulation, which caused significant heat transfer in it through the gas and along the glued layers of SVTI. The effective coefficient of thermal conductivity $\lambda_{ef}$ for such thermal insulation on a serial cryocontainer turned out to be equal to $28.8 \cdot 10^{-5}$ W / (m · K). This is 9.3 times higher than the value of this parameter for a calorimetric sample of this heat insulation.

Due to the absence of other domestic technologically advanced cushioning insulating materials for use in SVTI to reduce the content of the adhesive base, the SVTI–7 glass veil was pretreated in a special chamber by heating to 370–380 K with evacuation. As a result, the layers of the SVTI package on the cryocontainer did not stick together, and the thermal conductivity of their thermal insulation decreased to $14.1 \cdot 10^{-5}$ W / (m · K) (~ 2 times). At the same time, the service life ($R$) of a cryocontainer with liquid nitrogen (with one-time filling) with the use of improved thermal insulation increased from 75–80 days to 150–155 days (1.9–2 times).

In [17], it was also found that during long-term (for up to 15 years) operation of cryocontainers with liquid nitrogen, their thermal characteristics deteriorate by 42–85%. The reason for this turned out to be cryocondensate, which was formed on the cold layers of thermal insulation from the gas separation products pumped through them. This caused an increase in the degree of emissivity of the thermal insulation layers, their temperature and radiant heat transfer.

In this regard, a method was proposed for protecting the cold areas of the SVTI from the "poisonous" action of the resulting cryocondensate, consisting of gas separation products. The method was based on changing the direction of their pumping in the opposite direction (to the warm wall of the cryocontainer) through the optimal number (35) of the outer perforated layers of the SVTI package. Then these gases were evacuated using a loop-shaped evacuation process, which was first identified in the thermal insulation of cryocontainers [17]. The use of such, for the first time, the design of the SVTI package on new cryocontainers, allows for a long-term preservation of their initial thermal characteristics, which, during operation, gives a significant economic effect [17].

It is known from the literature [18] that the effective heat transfer through the SVTI package on the cryocontainer ($\lambda_{ef}$) can be carried out by the radiant ($\lambda_{rad}$), contact-conductive ($\lambda_{k.k}$) and gas ($\lambda_g$) components:

$$\lambda_{ef} = \lambda_{rad} + \lambda_{k.k} + \lambda_g. \tag{1}$$

According to theoretical as well as experimental data obtained on a calorimeter, it has been established that when the optimum pressure $P_o \leq 10^{-3}$ Pa in the thermal insulation is reached, the heat transfer through the gas becomes insignificant [18]. Therefore, the next task to create a scientific and technical basis for the calculation, design and manufacture of highly effective thermal protection for cryocontainers was the development of a technology for their manufacture with an optimal vacuum in SVTI.

Below are the techniques that were used in our development.

## 2. Research methods

The studies were carried out (as in [17]) on cryocontainers X – 34B with a capacity of 34 liters. As thermal insulation, we used an SVTI composition made of a screen polyethylene terephthalate film with double-sided aluminization (PET–DA) and interlining glass veil SVTI–7. This glass veil, according to the method described above, was subjected to preliminary thermal vacuum degassing in a special chamber at a temperature of 370–380 K with evacuation to $10^{-2}$ Pa for 8 hours to remove the PVA glue base from its structure to prevent adhesion of the SVTI layers.

The objective of these studies was to develop a technological process of thermal vacuum degassing for the insulating cavities of cryocontainers in electric furnaces, which would make it possible to obtain an optimal vacuum $P_o \leq 10^{-3}$ Pa during their subsequent filling with liquid nitrogen.

The external view of the experimental cryocontainer, assembled with a docking armature before being placed in an electric furnace, is shown in Fig. 1.

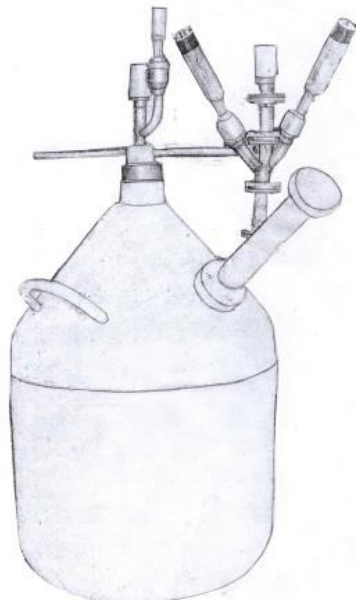

*Fig. 1*. Appearance of a cryocontainer with docking fittings before installation in an electric furnace.

In Fig. 2 shows an electric furnace with an experimental cryocontainer and an evacuation system used.

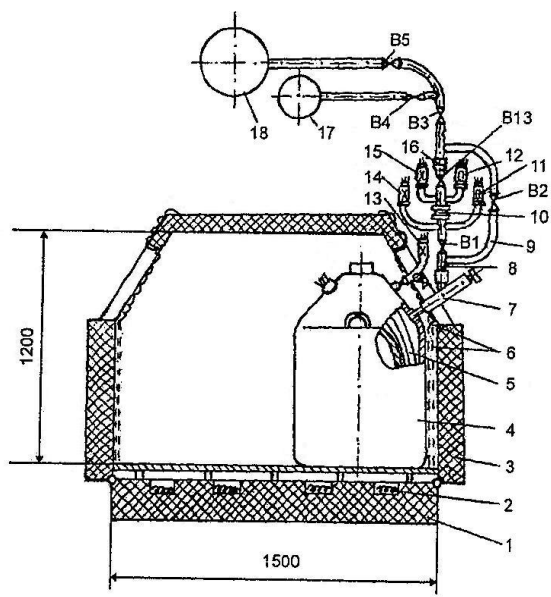

*Fig. 2*. Electric furnace with an experimental cryocontainer.
*1* – thermal insulation of an electric furnace; *2* – heating coil; *3* – electric furnace loading door; *4* – cryocontainer; *5* – layers of SVTI; *6* – curtains to reduce heat loss; *7* – docking device for connecting the evacuation nozzle of the cryocontainer with the evacuation system; *8, 9* – vacuum lines; *10* – diaphragm with calibration hole; *11–15* – manometric pressure sensors type PMT–2, PMI–2 and PMI–3–2; *16* – bellows; *17, 18* – collectors for rough and fine pumping, respectively.

The pressure in the insulating cavity of the cryocontainer under study is determined using pressure sensors PMT–2, PMI–2 and PMI–3–2 with an error of ± 15%. The temperature distribution over the average integral thickness of the thermal insulation in the cryocontainer [17] was measured by junction of calibrated copper – constantan thermocouples with an error of ± 0.08 K.

After thermal vacuum degassing in an electric furnace, the cryocontainers were subjected to thermal tests with liquid nitrogen according to the procedure described in [17]. For them, the total heat gain was determined, as well as for individual structural elements.

This made it possible to calculate their effective thermal conductivity coefficient (with an error of 3–5%) by the value of the heat flux through the thermal insulation ($Q_{ms}$) according to the equation [18]:

$$\lambda_{ef} = \frac{Q_{ms} \cdot \delta_{ms.m}}{F_m \cdot \Delta T_m}, \tag{2}$$

where $\delta_{ms.m}$ – the average integral thickness of the SVTI package over the entire surface of the inter-wall cavity of the cryocontainer X–34B (0.071 m);

$F_m$ – the average surface of the layers of thermal insulation on the cryocontainer X–34B (0.86 m$^2$);

$\Delta T_m$ – temperature difference between the outer and inner surfaces of a cryocontainer filled with liquid nitrogen (219 K).

### 3. Research results

In the experiments carried out, at certain time intervals ($\tau$), the attainable pressure was determined in the SVTI layers of the evacuated cryocontainer (placed in an electric furnace at a temperature of 390 K). After that, the cryocontainer was disconnected from the vacuum system of the electric furnace. Its evacuation nipple was hermetically sealed. The cryocontainer was removed from the electric furnace. After 4–5 hours, thermal tests with liquid nitrogen were carried out for it. Their goal was to determine the achieved value of the effective coefficient of thermal conductivity $\lambda_{ef}$ for the SVTI experimental cryocontainer at a steady pressure.

During such tests, 2–3 cryocontainers (out of 10) developed cracks in the glued joint of the fiberglass nozzle with the inner container. As a result, these cryocontainers suffered from leakage and they were rejected. To prevent this, the manufactured hot cryocontainers for cooling began to be held for 2 days under ambient conditions. Only after that did they begin to undergo tests with liquid nitrogen.

For the first time, the change in pressure in the thermal insulation of a cryocontainer during this cooling process was investigated. The results of such studies on the example of cryocontainer *N*1 (after 100 hours of evacuation in an electric furnace), as well as for thermal insulation of cryocontainer *N*2 (after 150 hours of such evacuation) are shown in Fig. 3 by the dependences $P_{ins}(\tau)$ 3 and 5, respectively.

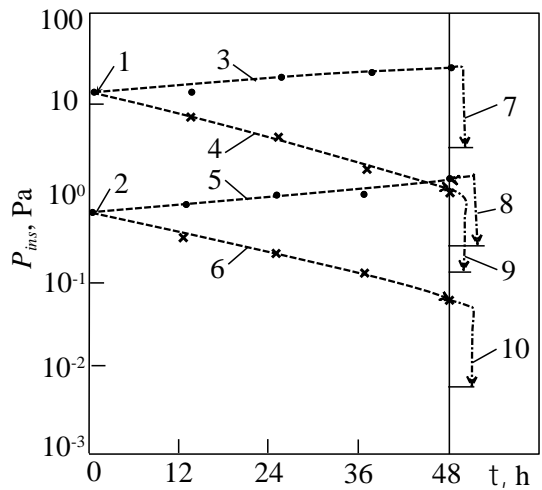

*Fig. 3*. Changes in pressure in the insulating cavity of cryocontainers during various cooling processes in ambient conditions.

*1, 2* – pressure in the SVTI layers of hot cryocontainers *N*1 and *N*2, respectively; *3, 5* – pressure change during cooling under ambient conditions for thermal insulation of cryocontainers *N*1 and *N*2, respectively; *4, 6* – change in pressure in the layers of SVTI when they are pumped out by a turbomolecular pump with cooling under ambient conditions for experimental cryocontainers *N*1 and *N*2, respectively; *7–10* – trajectories of pressure decrease in the insulating cavity of cryocontainers after filling with liquid nitrogen.

The analysis of the obtained results shows that when the cryocontainers are cooled at ambient temperature, a gradual increase in pressure occurs in their thermal insulation. At the same time, in cryocontainer *N*1, this parameter from 15 Pa (obtained in it in an electric furnace after 100 hours of evacuation) increases after 48 hours to 63 Pa (by 4.2 times). For cryocontainer *N*2, the pressure in the thermal insulation during the same time increased by a smaller value, from 0.85 Pa to 2.6 Pa (by ~ 3 times). The reason for this was the longer (by 50 hours) pumping out of the materials of the insulating cavity in this cryocontainer in comparison with the cryocontainer *N*1. As a result, a smaller volume of dissolved gases remained in the material structure of the *N*2 cryocontainer.

After the end of cooling under ambient conditions, the cryocontainers were filled with liquid nitrogen. As a result, the pressure in the thermal insulation of cryocontainer *N*1 decreased (trajectory 7) from 63 Pa to 12 Pa (by 5.3 times). In this case, the pressure for the SVTI cryocontainer *N*2 decreased (trajectory 8) from 2.6 Pa to 0.38 Pa (by 6.8 times).

The results obtained showed the ineffectiveness of using the process of cooling hot cryocontainers at ambient temperature before testing with liquid nitrogen. Therefore, it was proposed to carry out the cooling of cryocontainers after an electric furnace with their simultaneous evacuation with a turbomolecular pump.

The results of such studies are shown in *Fig. 3* by the dependences $P_{ins}(\tau)$ *4* and *6*. It follows from them that during this process the pressure in the layers of the SVTI cryocontainer *N*1 decreases from 15 Pa to 2.2 Pa by 6.8 times. After the subsequent filling of the cryocontainer *N*1 with liquid nitrogen, the pressure in its SVTI decreases (according to trajectory *9*) to 0.3 Pa. For cryocontainer *N*2, after evacuation of the SVTI by a turbomolecular pump, the pressure decreases from 0.85 Pa to $8 \cdot 10^{-2}$ Pa (by 9.6 times). After its subsequent filling with liquid nitrogen, the pressure decreases (trajectory *10*) to $8.5 \cdot 10^{-3}$ Pa (by ~ 9 times). Thus, the optimal pressure $P_0 \leq 10^{-3}$ Pa in the cryocontainers has not yet been reached.

In this work, similar studies were carried out for the thermal insulation of cryocontainers (when they are heated in an electric furnace) in the range from 50 hours to 275 hours. The results obtained are shown in *Fig. 4* by the dependence $P_{ins}(\tau)$ *1*.

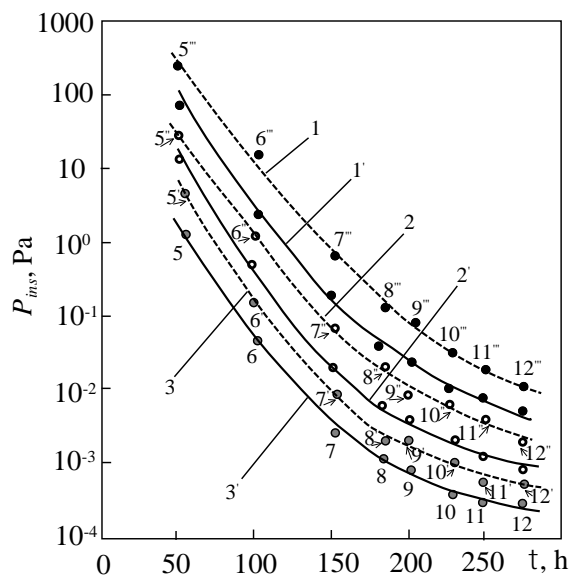

*Fig. 4.* Change in pressure in the insulating cavity of the cryocontainer.

*1, 1'* –in the process of evacuating cryocontainers in an electric furnace at a temperature of 390 K using heat-insulating materials unmodified and modified by $N_2$ molecules, respectively; *2, 2'* – after evacuation by a turbomolecular pump of cryocontainers (removed from the furnace) with unmodified and modified $N_2$ molecules, insulating materials, respectively; *3, 3'* – after filling the cryocontainers with liquid nitrogen, evacuated by a turbomolecular pump with unmodified and modified $N_2$ molecules, insulating materials, respectively.

Changes in the pressure in the insulating cavity of these cryocontainers after their evacuation for 48 hours with a turbomolecular pump in the cooling mode, as well as subsequent filling with liquid nitrogen are shown in *Fig. 4* by the dependences $P_{ins}(\tau)$ *2* and *3*, respectively.

Analysis of the dependence $P_{ins}(\tau)$ *3* in *Fig. 4* shows that after 7 days of evacuation of the insulating cavity of cryocontainer *N8'* (according to the technological regulations for their manufacture in an electric furnace), subsequent evacuation with a turbomolecular pump for 48 hours and filling with liquid nitrogen, the optimal vacuum is not yet reached in it. The insulating cavity was pumped out only to $7 \cdot 10^{-3}$ Pa. Further studies have shown that the optimal pressure can be obtained only after pumping out the insulating cavity for 225 hours or more in an electric furnace, followed by evacuation with a turbomolecular pump and filling with liquid nitrogen. This can be seen from the dependence $P_{ins}(\tau)$ *3*.

In this study, for the thermal insulation of each cryocontainer after tests with liquid nitrogen, their effective thermal conductivity coefficients $\lambda_{ef}$ were determined according to equation (2), which are shown in Fig. 5 are represented by the dependence $\lambda_{ef}(P_{ins})$ *1*. The results obtained refer to thermal insulation on similarly designated cryocontainers in *Fig. 4*.

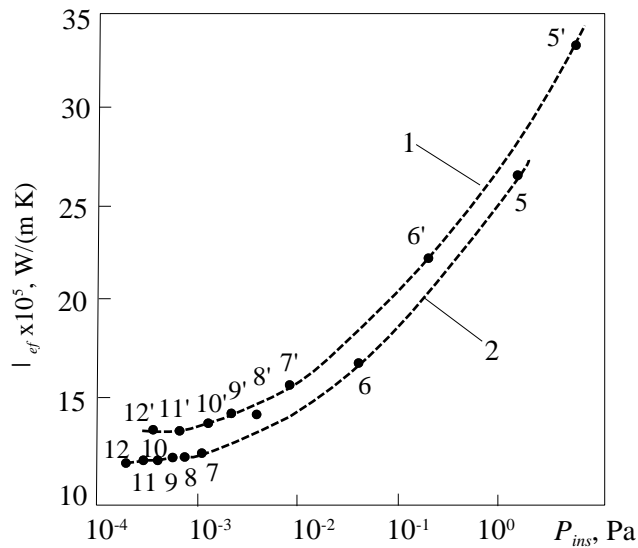

*Fig. 5.* Dependence of the effective coefficient of thermal conductivity for thermal insulation on nitrogen cryocontainers on pressure.

*1* – in cryocontainers with thermal insulation PET – DA + SVTI – 7 (the SVTI – 7 glass veil and the adsorbent of the vacuum pump were subjected to preliminary thermal vacuum degassing at 370–380 K in a separate chamber for 8 hours); *2* – in cryocontainers with similar thermal insulation (the SVTI–7 glass veil and the adsorbent of the vacuum pump were subjected to the same preliminary treatment for 12 hours, as well as their subsequent modification with dry nitrogen molecules).

The analysis of the results shows that with a decrease in pressure in the layers of the SVTI cryocontainers (due to an increase in the duration of their evacuation (*Fig. 4*)), the effective coefficient of thermal conductivity decreases sharply. For example, after 50 hours of evacuation of a hot cryocontainer *5'* (*Fig. 4*) and obtaining (as a result of subsequent cooling with liquid nitrogen) a pressure of ~ 8 Pa in it (*Fig. 5*), the effective thermal conductivity coefficient of its SVTI layers becomes equal to $35 \cdot 10^{-5}$ W / (m · K). When evacuating the thermal insulation for a longer time (for example, regulatory), equal to 168 hours, the optimal pressure in it is not yet reached (and is equal to $7 \cdot 10^{-3}$ Pa). In this case, the effective coefficient of thermal conductivity decreases to $15 \cdot 10^{-5}$ W / (m · K). After evacuation for 225 hours, the pressure in the thermal insulation of the *N*10' cryocontainer has already decreased to the optimal value $P_0 = 10^{-3}$ Pa, and the effective thermal conductivity coefficient decreases to $14.1 \cdot 10^{-5}$ W / (m · K).

With further evacuation of the thermal insulation of cryocontainers *N*11' and *N*12' for 250 and 275 hours to a pressure of $7 \cdot 10^{-4}$ Pa and $3 \cdot 10^{-4}$ Pa, their effective thermal conductivity coefficients turned out to be practically unchanged (equal to $14.1 \cdot 10^{-5}$ W / (m · K) and $14.3 \cdot 10^{-5}$ W / (m · K)).

The results obtained made it possible to experimentally establish for the first time that in the thermal insulation of cryocontainers with a significant number (80–100) of the SVTI layers and an adsorption vacuum pump on the cold wall, when the pressure reaches below $10^{-3}$ Pa, the effective thermal conductivity coefficients are remain practically unchanged. In this regard, pressures below $10^{-3}$ Pa can also be considered optimal for the heat-shielding layers of SVTI on cryocontainers.

Thus, the studies carried out made it possible to establish that after carrying out the process of thermal vacuum degassing in an electric furnace for the cryocontainers for a regulated time of 7 days (168 hours), its further evacuation for 48 hours with a turbomolecular pump and subsequent filling with liquid nitrogen), the optimum pressure $P_0 \leq 10^{-3}$ Pa in its thermal insulation has not yet been reached. The use of pumping fittings of a larger flow area (in comparison with the used diameter of 0.018 m) on cryocontainers to accelerate the evacuation process was also impossible, due to the complexity of their subsequent reliable sealing with special plugs.

In this regard, a developed new technology was proposed to solve the problem of accelerating the process of evacuating the insulating cavity of a cryocontainer. Its essence is as follows. It was shown in [19] that water molecules are the main gas separation products for heat-insulating materials. The reason for this is their high (by 3–4 times) energy of adsorption in comparison with other gases in the atmosphere [20, 21]. Therefore, they are mainly contained on the surface and in the structure of all materials in the environment. In this regard, it was proposed to speed up the evacuation process for the main materials used in the thermal insulation of cry-ocontainers (SKT–4 coal and SVTI–7 cushion glass veil) their preliminary processing in order to replace water molecules in their structure with molecules of less active dry nitrogen (or air). To implement the proposed technology, a special vacuum chamber was manufactured (using a pumping nozzle with a flow area increased to 0.05 m). The chamber was heated after filling with insulating materials to 370–380 K and evacuated to $10^{-2}$ Pa for 12 hours. After that, it was filled with dry nitrogen. Prepared in this way, modified with nitrogen, the materials were stored for 8–16 hours in a special hermetic container (filled with $N_2$) in an electric furnace at a temperature of ~ 370 K before being used in cryocontainers.

The process of pumping out in an electric furnace of cryocontainers with heat-insulating materials modified with $N_2$ molecules is represented by the dependence $P_{ins}(\tau)$ *1'* in *Fig. 4*. It follows from it that after saturation with $N_2$, the SVTI layers are evacuated 7–8 times faster in comparison with the thermal insulation not subjected to modification (experimental dependence *1* in *Fig. 4*). The process of evacuation of the insulating cavities of cryocontainers with materials modified with $N_2$ molecules and their evacuation by a turbo-molecular pump is also accelerated, which follows from a comparison of the dependences $P_{ins}(\tau)$ *2'* and *2*.

The change in pressure in the interstitial cavity of cryocontainers with materials modified with $N_2$ molecules after filling them with liquid nitrogen is shown in *Fig. 4* by the dependence $P_{ins}(\tau)$ *3*.

For the modified insulating materials used on cryocontainers, the dependences of their effective thermal conductivity coefficients on pressure were also determined, which are shown in *Fig. 5* by the dependence $\lambda_{ef}(P_{ins})$ *2*. The achieved improvement in the thermal characteristics of insulating materials due to their modification by $N_2$ molecules is clearly seen from the comparison of the effective thermal conductivity coefficients for the layers of the SVTI cryocontainers on the dependence $\lambda_{ef}(P_{ins})$ *2*, as well as on the dependence $\lambda_{ef}(P_{ins})$ *1* with unmodified materials on cryocontainers. It follows from this that after vacuumization for 50 hours, the thermal conductivity coefficient for unmodified layers of SVTI on cryocontainer *N5'* becomes equal to $34 \cdot 10^{-5}$ W / (m · K), and for modified thermal insulation on cryocontainer *N5* for the same pumping time $\lambda_{ef}$ decreases to $27 \cdot 10^{-3}$ W / (m · K) (by 26%). After 100 hours of evacuation, the thermal conductivity $\lambda_{ef}$ for the mounted thermal insulation in the cryocontainer *N6'* turned out to be equal to $22 \cdot 10^{-5}$ W / (m · K), and for the thermal insulation modified by $N_2$ for the cryocontainer *N6*, the value of this parameter decreases to $18 \cdot 10^{-5}$ W / ( m · K) (by 22%). A similar comparison for thermal insulation on cryocontainer *N7'* and *N7* showed a decrease in the difference between their thermal characteristics by a smaller value equal to 17%.

The established more significant decrease in pressure and the values of effective thermal conductivity coefficients for thermal insulation materials modified with $N_2$ molecules is caused by a significantly lower adsorption energy of $N_2$ in comparison with water molecules contained in unmodified thermal insulation of cryocontainers on the dependence $\lambda_{ef}(P_{ins})$ *1*. At the same time, the revealed increased difference between the compared values of the thermal conductivity coefficients at the beginning of the evacuation process is explained by the more intensive pumping out of gas molecules from the layers adsorbed on the surface of the materials, where they have the lowest binding energies. With an increase in the time of the evacuation process, the differences between the compared coefficients of thermal conductivity decrease. This is due to the transition of the process of evacuation of gases from the more developed internal structure of the material, where they have a significantly higher adsorption energy in comparison with surface adsorption layers.

These studies also made it possible to establish that for cryocontainers with thermal insulating materials modified with $N_2$ molecules, after 150 hours of evacuation in an electric furnace (followed by evacuation with a turbomolecular pump and filling with liquid nitrogen), the optimal pressure $P_0 = 10^{-3}$ Pa (cryocontainer *N7* in *Fig. 5*). The effective coefficient of thermal conductivity for the insulating material modified by $N_2$ molecules on the *N7* cryocontainer becomes equal to $12.4 \cdot 10^{-5}$ W / (m · K). When the pressure drops below $10^{-3}$ Pa, the thermal conductivity for similar materials on cryocontainers *N8 – N12* becomes (as can be seen from *Fig. 5*) practically unchanged, as for unmodified insulating materials on cryocontainers *N10 – N12*. The thermal conductivity coefficients obtained at an optimal pressure $P_0 \leq 10^{-3}$ Pa for thermal protection materials on cryocontainers *N8–N12* turned out to be ~ 18% lower in comparison with the value of this parameter for SVTI on cryocontainer *N8'* after 168 hours of evacuation, according to the routine technology for their serial production. Thus, a significant improvement in the quality of the manufactured cryocontainers has been achieved. In this regard, it became possible to increase productivity for electric furnaces. In addition, from the results of the study, it was also established that the effective coefficients of thermal conductivity for the SVTI layers on cryocontainers *N8–N12* equal to $(12.2–12.4) \cdot 10^{-5}$ W / (m · K) turned out to be ~ 11% lower in comparison with analogous parameters for the layers of SVTI on cryocontainers *N10'–N12'* with a value of $(14.1–14.3) \cdot 10^{-5}$ W / (m · K). The achieved decrease in the $\lambda_{ef}$ values for the SVTI materials on the *N7–N12* cryocontainers could not result from their modification with $N_2$ molecules. The use of such materials in the inter-wall cavity of these cryocontainers only contributes to the acceleration of the process of their evacuation in comparison with insulating unmodified materials on cryocontainers *N5'–N12'*. The established improvement (by 11%) of thermal characteristics for thermal insulations on cryocontainers *N5–N12* occurred as a result of the use of an SVTI–7 glass veil with a new, more efficient structure in their layers. An improvement in the structure of this material was obtained as a result of a significantly longer (for 12 hours) preliminary evacuation of it in a special chamber at a temperature of 380 K, which contributed to a decrease in the contact-conductive component $\lambda_{k.k}$ of heat transfer in SVTI. To insulate the *N5'–N12'* cryocontainers, the glass veil was degassed for only 8 hours. It follows from this that an improvement in the thermal characteristics of thermal insulation can also be obtained by using in their structure lining materials with optimal thermophysical, optical, vacuum and other characteristics. The development of such cushioning materials can be the topic of special research in the direction of creating highly efficient thermal insulation with a thermal conductivity coefficient close to its minimum value on a calorimeter.

Further, it was of interest to evaluate the contribution of the gas component to the effective thermal conductivity from the obtained experimental data with increasing pressure in the thermal insulation. It was shown above that at a pressure of $P_0 \leq 10^{-3}$ Pa, the effective thermal conductivity does not practically depend on pressure and is determined by the contribution of radiant $\lambda_{rad}$ and contact-conductive $\lambda_{k.k}$ components of heat transfer. At pressures greater than the optimal value, an increase in the effective thermal conductivity coefficients $\lambda_{ef}$ is caused by an increase in the gas component of heat transfer $\lambda_g$. It follows from this that the contribution of this component to the effective thermal conductivity can be determined from the ratio:

$$\lambda_g = \frac{\lambda_{ef} - (\lambda_{rad} + \lambda_{k.k})}{\lambda_{ef}} \cdot 100, \% \qquad (3)$$

This calculation was carried out using the experimental results obtained for the SVTI layers modified with $N_2$ molecules and shown in *Fig. 5* by the dependence $\lambda_{ef}(P_{ins})$ *2*. For it, the joint radiant and contact-conductive heat transfer $(\lambda_{rad} + \lambda_{k.k})$ at pressure $P_0 = 10^{-3}$ Pa was determined above and amounts to $14.2 \cdot 10^{-5}$ W / (m · K). The results of the study are presented in *Fig. 6* by the dependence $\lambda_g(P_{ins})$.

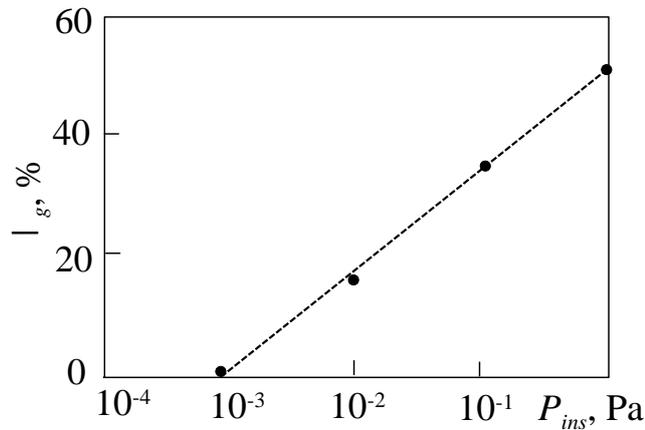

*Fig. 6*. Increase in the contribution of the gaseous component of heat transfer ($\lambda_g$) to the effective thermal conductivity ($\lambda_{ef}$) with increasing pressure in the heat-insulating composition of PET–DA+SVTI–7 layers on a cryocontainer with liquid nitrogen.

From the dependence $\lambda_{ef}(P_{ins})$ *2* in *Fig. 5*, as well as the results in *Fig. 6* it follows that with an increase in pressure in the layers of thermal insulation, the effective coefficient of thermal conductivity increases sharply from the gas component. At the same time, for example, for the thermal insulation of a cryocontainer *N*5 with a thermal conductivity coefficient $\lambda_{ef} = 27 \cdot 10^{-5}$ W / (m · K) and a pressure in the SVTI layers equal to 1 Pa, the contribution of the gas component to the effective thermal conductivity is 54%.

Further, for the SVTI package on an industrial cryocontainer with optimal pressure, in order to study the features of the mechanisms of heat transfer, the temperature dependences of the effective thermal conductivity and its components were first investigated for thermal insulation on the cryocontainer, according to the relation [18]:

$$\lambda_{ef}(T) = \lambda_{rad}(T) + \lambda_{k.k}(T). \tag{4}$$

To carry out these studies, an experimental cryocontainer was made with the optimal pressure in the SVTI layers, thermal conductivity of $12.4 \cdot 10^{-5}$ W / (m · K) and temperature sensors (copper-constantan calibrated thermocouples) according to their mean integral thickness ($\delta_{m.ins}$), according to the described technology [17]. The temperature profile $T(x/\delta)$ measured in this way for a given thermal insulation is shown in *Fig. 7*.

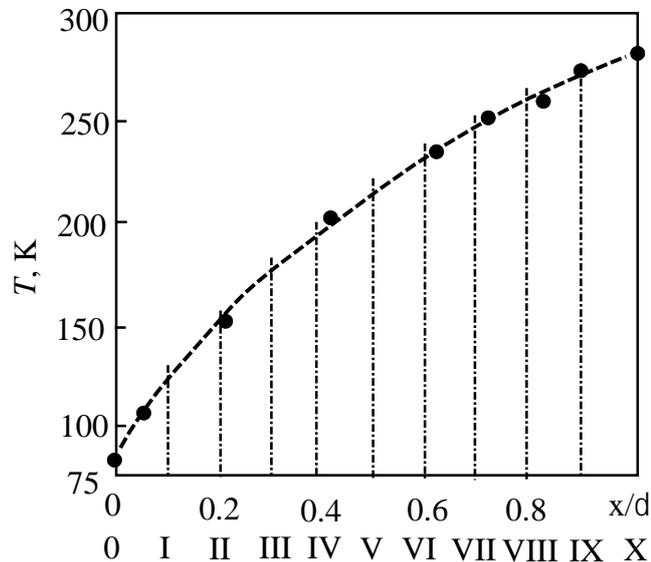

*Fig. 7*. Temperature distribution over the mean integral reduced thickness ($x/\delta$) of the SVTI layers in an industrial cryocontainer at boundary temperatures (78÷297) K.
I–X – conditional breakdown of the temperature profile $T(x/\delta)$ into elementary sections.

This process is evidently monotonic (without inflection points), which indicates the absence in the structure of the thermal insulation package of a local area with increased gas pres-sure from the products of gas separation of the insulating material [17]. Further, the cross-section of the thermal insulation, where the thermocouples are located, is conventionally divided into elementary sections $\Delta\delta$ (as shown in *Fig. 7*). Then the thermal conductivity coefficient for any $i$-th section of the SVTI can be determined by the Fourier equation [18]:

$$\lambda_{ef\,i} = q \cdot \frac{\Delta\delta}{\Delta T_i}, \qquad (5)$$

where $\Delta T_i$ – temperature drop on the $i$-th section of thermal insulation;

$q$ – specific heat flux (for each $i$-th section of the SVTI package is taken equal to its average value for thermal insulation).

The error in determining the coefficient $\lambda_{ef}(T)$ is (3÷5)%. A certain temperature dependence for the effective thermal conductivity of the thermal insulation of the experimental cryocontainer is shown in *Fig. 8* by the dependence $\lambda_{ef}(T)$ *1*.

Radiant heat transfer in the $i$-th section of the SVTI package is calculated by the equation [18]:

$$\lambda_{rad\,i}(T) = 4\frac{\varepsilon}{2-\varepsilon}\frac{\delta}{N}\sigma T_m^3, \qquad (6)$$

where $\varepsilon$ – the degree of emissivity of thermal insulation screens was taken from the data [22];

$N$ – number of layers of thermal insulation in the cryocontainer;

$\sigma$ – Stefan – Boltzmann constant;

$T_m$ – average temperature of thermal insulation in the $i$-th section.

The calculated dependence $\lambda_{rad\,i}(T)$ *3* for the thermal insulation of the experimental cryocontainer is shown in *Fig. 8*.

Since the pressure in the thermal insulation of the experimental cryocontainer was optimal, and heat transfer is carried out only by the radiant and contact-conductive components of the effective thermal conductivity, the temperature dependence for the contact-conductive thermal conductivity was determined from the relation:

$$\lambda_{k.k\,i}(T) = \lambda_{ef\,i}(T) - \lambda_{rad\,i}(T).$$

The calculated results are shown in *Fig. 8* by the dependence $\lambda_{k.k}(T)$ *2*. The dependences $\lambda_{ef}(T)$ *1*, $\lambda_{k.k}(T)$ *2* and $\lambda_{rad}(T)$ *3* were obtained for the first time using the developed method for a heat-shielding package on a cryocontainer with liquid nitrogen.

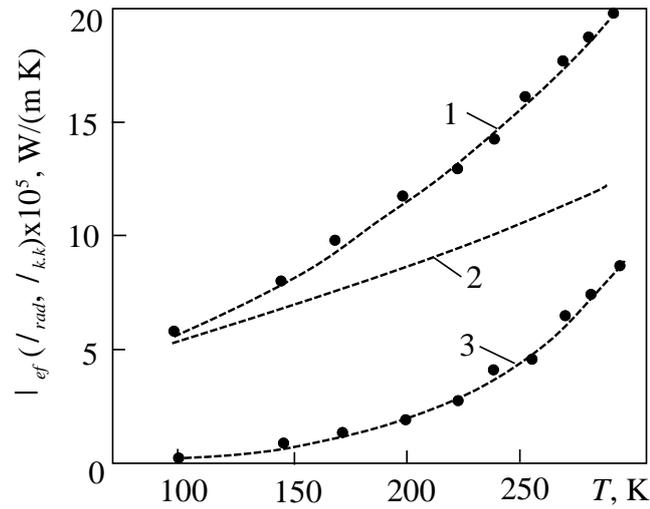

*Fig. 8.* Change with temperature of the effective thermal conductivity coefficient of SVTI and its components on a cryocontainer with liquid nitrogen.

*1* – effective thermal conductivity $\lambda_{ef}$; *2* – contact-conductive thermal conductivity $\lambda_{k.k}$; *3* – radiant thermal conductivity $\lambda_{rad}$.

They, as you can see, change with different intensity with increasing temperature. At the same time, it was found that the contribution to the effective thermal conductivity $\lambda_{ef}$ of the contact-conductive component $\lambda_{k.k}$ (equal to $9.4 \cdot 10^{-5}$ W / (m · K)) is 67%, and the radiant $\lambda_{rad}$ (value $4.7 \cdot 10^{-5}$ W / (m · K)) – 33%. From the results obtained, it was concluded that the layers of the SVTI on the investigated cryocontainer during machine insulation were mounted with non-optimal assembly tension forces. This was the reason for obtaining in the SHTI increased (by ~ 2 times) contact-conductive heat transfer $\lambda_{k.k}$ in comparison with radiant $\lambda_{rad}$.

There were no methods and devices for controlling the tensile forces for the SVTI strips mounted on cryocontainers when they were isolated. The mounting parameters, which determine the technology of machine accelerated isolation of cryocontainers with narrow strips of SVTI, also remained unknown.

Therefore, the next task for the development of a technology for the machine-made production of highly efficient SVTI packages on cryocontainers should be to identify all the parameters of machine isolation, develop methods and tools for their measurement and control, as well as determine the optimal values.

**Conclusions**

1. A technology has been proposed and developed to improve the thermal characteristics of the manufactured cryocontainers by obtaining the optimal pressure $P_o \leq 10^{-3}$ Pa in their SVTI layers, at which the gaseous component of effective thermal conductivity is eliminated.

2. A technology has been developed to accelerate the process of evacuating the insulating cavities of cryocontainers by using insulating materials in them, from the structure of which water molecules were previously (in a separate chamber) pumped out and then replaced with dry nitrogen. As a result, the evacuation process is accelerated, because $N_2$ molecules have 3–4 times less adsorption energy in comparison with $H_2O$ molecules. This makes it possible to obtain the necessary optimal vacuum $P_o \leq 10^{-3}$ Pa in the inter-wall cavity of the cryocontainers for 150 hours of evacuation in an electric furnace, which is ~ 20 hours faster in comparison with the scheduled time.

3. When the optimal pressure of $10^{-3}$ Pa is reached, the effective thermal conductivity $\lambda_{ef}$ for the SVTI package on the cryocontainer becomes equal to $14.2 \cdot 10^{-5}$ W / (m · K). This is 11% lower than the previously achieved reduction of this parameter to $12.1 \cdot 10^{-5}$ W / (m · K) as a result of the elimination of adhesion of the SVTI layers on cryocontainers.

4. With a decrease in pressure below $10^{-3}$ Pa, the effective thermal conductivity $\lambda_{ef}$ (for example, 4 cryocontainers with such thermal insulation) remained practically unchanged, equal to $(14.1–14.3) \cdot 10^{-5}$ W / (m · K). Thus, for the first time for thermal insulation on a cryocontainer, the possibility of achieving an optimal vacuum is shown, at which the gas component of heat transfer is excluded, and the effective thermal conductivity is determined only by contact-conductive and radiant thermal conductivity.

5. A technique has been developed for determining the contribution to the effective thermal conductivity for the SVTI on a cryocontainer sel of its components from the temperature profile measured for it. For thermal insulation on a cryocontainer with optimal pressure $P_o \leq 10^{-3}$ Pa and thermal conductivity (for example) $14.1 \cdot 10^{-5}$ W / (m · K), the contribution of the radiant ($\lambda_{rad}$) component of heat transfer ($4.7 \cdot 10^{-5}$ W / (m · K)) is 33%, and the contact-conductive ($\lambda_{k.k}$) component ($9.5 \cdot 10^{-5}$ W / (m · K)) – 67%.